\documentstyle[preprint,aps]{revtex}
\newcommand {\tr}{\mbox{ tr }}
\newcommand{\eq}[1]{eq.~(\ref{#1})}
\newcommand{\la}[1]{\label{#1}}
\newcommand{\beq}{\begin{equation}}
\newcommand{\eeq}{\end{equation}}
\newcommand {\bear}{\begin{eqnarray}}
\newcommand {\ear}{\end{eqnarray}}

\def\theequation{\arabic{section}.\arabic{equation}}
\def\appendix{\par
 \setcounter{section}{0}
 \def\thesection{Appendix}
 \def\theequation{\Alph{section}.\arabic{equation}}}

\begin{document}

\preprint{PNPI-TH-2068}
\title{ Effective Chiral Lagrangian from Dual Resonance Models }
\author{
Maxim V. Polyakov$^1$\footnote{e-mail: maxpol@thd.pnpi.spb.ru}
 and Vladimir V. Vereshagin$^2$\footnote{e-mail:
 vereshagin@phim1.niif.spb.su} }
 \address{$^1$Petersburg Nuclear  Physics Institute, Gatchina
 188350, St. Petersburg , Russia \\ $^2$ Institute of Physics,
 St. Petersburg University, 198904 St.  Petersburg, Russia }
 \date{ September 1995 }
 \maketitle

\begin{abstract}
Parameters of the effective chiral lagrangian (EChL) of orders
$O(p^4)$   and $O(p^6)$  are extracted from low--energy
behaviour of  dual resonance models for $ \pi\pi$ and
$\pi K$ scattering amplitudes.
Dual resonance models are considered to be  good candidates for
the resonance spectrum and for  hadronic scattering amplitudes
in the large $N_c$ limit of  QCD. We discuss  dual
resonance models in the presence of spontaneous and explicit
chiral symmetry breaking.  Obtained parameters of the EChL are
used to estimate chiral
corrections up to the sixth order to various low--energy
characteristics of $\pi\pi$ and $\pi K$ scattering amplitudes.

\end{abstract}

\newpage
\noindent
\section{Introduction}
\setcounter{equation}{0}

The technique of the Effective Chiral Lagrangian (EChL)
provides us with a systematic way of low-energy expansion of
correlators of different colourless currents in  Quantum
Chromodynamics \cite{WeiOld,Wei,GL}. The information about
large distance behaviour of the QCD  is hidden in a finite set
of coupling constants if we restrict ourselves to finite order
in the momentum expansion.  In the language of the EChL the
expansion of the Green functions in external momenta and quark
masses corresponds to an expansion in number of meson loops.
With increasing order of the chiral expansion one has to extend
the EChL by introducing couplings with increasing number of
derivatives and increasing power of quark masses.  Number of
terms of the EChL grows rapidly with the expansion order.   For
example, in the leading order there are two possible chiral
couplings, in the next-to-leading order Gasser and Leutwyler
\cite{GLsu3} determined ten low--energy constants which are not
fixed by chiral symmetry requirements.  At the
next-to-next-to-leading ($O(p^6)$) order there are more than a
hundred new low--energy constants \cite{feasch}. An estimate of
these sixth order constants is important to calculate the
analytical (polynomial) part of the chiral sixth order
corrections to the meson Green functions. Such kind of
corrections arise from tree graphs of the sixth order EChL
(${\cal L}^{(6)}$), other contributions in this order of the
chiral counting arise from one and two loops graphs of the
fourth (${\cal L}^{(4)}$)  and second order EChL (${\cal
L}^{(2)}$) correspondingly.  The loop corrections are expressed
in terms of the known parameters of the fourth and second order
EChL, whereas the polynomial corrections being technically
easily calculable (tree graphs) depend on unknown constants.
Hence to calculate the complete next-to-next-to-leading
contributions to different meson Green functions one has to pin
down the sixth order EChL coupling constants.

In this paper we determine some of the coupling constants of
the sixth order EChL, in particular we calculate the polynomial
part of the sixth order corrections to the elastic $\pi \pi$
and $\pi K$ scattering amplitudes.  To this end we  use large
$N_c$ (number of colours) arguments or equivalently impose the
Okubo--Zweig--Iizuki (OZI) rule. It is known that the fourth
order constants determined in ref.~\cite{GLsu3} respect the OZI
rule with good accuracy.  Moreover in the large $N_c$ limit the
parameters of the effective chiral lagrangian can be related to
the resonance spectrum
\cite{EckGasPicRaf,DonRamVal,BolPolVer,BolManPolVer} by
contraction of the resonance contributions to the
(pseudo)Goldstone scattering amplitudes.  To ensure the chiral
symmetry of resulting EChL one can either impose the chiral
symmetry on the coupling of resonances to pions
\cite{EckGasPicRaf,DonRamVal} or impose some relations on
resonance spectrum \cite{BolPolVer,BolManPolVer} in the spirit
of Weinberg's approach to the algebraic realization of the
chiral symmetry
\cite{Weinber_alg_real,Weinberg_mended_symmetry}. The latter
approach apart of predictions for the EChL parameters gives an
infinite set of equations for resonance spectrum. These
equations were derived and analyzed in ref.~
\cite{BolManPolVer}, it was shown there that the equations on
the spectrum of the $\pi\pi$ resonances ensure the duality
properties of the $\pi \pi$ scattering amplitude.
Phenomenologically the duality of hadronic amplitudes was
suggested in the sixties \cite{DolHorSch} as a certain relation
between two ways of describing scattering amplitudes: the Regge
pole exchange at high energies and resonance dominance at low
energies. Later explicit models for hadron interaction
implementing duality were constructed \cite{Veneziano} and
found to be in an agreement with experimental data
\cite{Frampton}.  Almost immediately it was found that the dual
resonance amplitudes arise naturally in the quantum theory of
the extended objects -- strings. Now there is considerable
theoretical belief that QCD in the large $N_c$ limit
corresponds to some string theory \cite{Polyakov}, though the
particular form of the theory is not found.  This task is
difficult because it involves comparing field theory (QCD) in
which we can not compute hadron amplitudes, with a string
theory in which basically all one can do is to compute $S$-
matrix in the narrow resonance approximation.  Manifestations
of possible underlying string dynamics in hadronic spectrum and
interactions were recently discussed in
refs.~\cite{Lew,Centr,mylec}.

 We make use of the dual resonance models (DRM) to
estimate of the parameters of the sixth order chiral
lagrangian. First, we study the conditions imposed by
low--energy theorems on the dual resonance (string) models.
Expand then the obtained amplitudes with `` built in ''
 soft--pion theorems at low energies and comparing the resulting
 expansions with those given by the EChL we are able to fix the
low--energy constants of the sixth order EChL.  In principle,
one can saturate sum rules relating the EChL and resonances
 spectrum derived in
\cite{EckGasPicRaf,DonRamVal,BolPolVer,BolManPolVer} by
phenomenological resonance spectrum, unfortunately
the corresponding sum rules for the sixth order EChL are very
sensitive to experimental uncertainties, especially in the
scalar channel where the spectroscopic data are controversial.
Instead of that, we shall use the dual resonance models as
models for resonance spectrum and their interactions.  These
models possess many attractive properties, in particular the
dual resonace amplitudes have a correct Regge high energy
behaviour and hence incorporate naturally the algebraic
realization of chiral symmetry \cite{Weinber_alg_real}.
Also they predict a correct resonance mass spectrum.

The paper is organized as follows. In sect.~\ref{echl_sect} we
introduce sixth order effective chiral lagrangian relevant for
our purposes and fix our notations. In sect.~\ref{ampl_sect}  we
discuss the polynomial part of the chiral corrections to the
$\pi\pi$ and $\pi K$ amplitudes in the next-to-next-to-leading
order. In particular we give an explicit expression for the
tree-level $\pi\pi$ and $\pi K$ amplitudes in the sixth order in
terms of the low--energy constants of the ${\cal L}^{(6)}$.
Dual resonance $\pi\pi$ and $\pi K$ amplitudes with
spontaneously and explicitly broken chiral symmetry are
constructed in sect.~\ref{dual_sect}. We show that the
soft--pion theorems impose very strong conditions on the dual
amplitudes; that enables us in sect.~\ref{low_dual_sect} to
calculate some of the low--energy constant of the sixth order
EChL and compare them with chiral quark model predictions.
Obtained parameters are used to estimate polynomial (analytical)
part of the sixth order contribution to the low--energy
scattering parameters (scattering lengths, slope parameters,
etc.). Our summary and conclusions are surveyed in
sect.~\ref{concl_sect}.

\section{ Effective Chiral Lagrangian to $O(p^6)$ }
\setcounter{equation}{0}
\label{echl_sect}

In the lowest order of momentum expansion $O(p^2)$ the interactions
of (pseudo)Goldstone mesons (pions, kaons and eta mesons) are
described by the famous Weinberg lagrangian \cite{WeiOld,Wei}:

\beq {\cal L}^{(2)}=\frac{F_0^2}{4}
tr(L_\mu L_\mu) + \frac{F_0^2 B_0}{4} tr(\chi),
\la{echl2}
 \eeq
where $\chi=2 B_0 (\hat{m}U+U^\dagger\hat{m} )$, $L_\mu=iU\partial_\mu
U^\dagger$,
$\hat{m}=\mbox{diag}(m,m,m_s) $ is a quark mass matrix and
$F_0$ and $B_0$ are low-energy coupling constants carrying
an information about long-distance behaviour of the QCD.
The latter are related to pion decay constant and quark
condensate in the chiral limit:
\bear
\nonumber
F_0 &=& \lim_{m_q \rightarrow 0}F_\pi, \\
\nonumber
B_0 &=& -\lim_{m_q \rightarrow 0} \frac{\langle \bar \psi \psi
\rangle}{F_\pi}.
\ear
The chiral field $U(x)$ is a unitary $3 \times 3$ matrix and
is parametrized in terms of eight pseudoscalar meson fields
$\pi$, $K$ and $\eta$:

\bear
\nonumber
U(x)&=& e^{i\Pi}, \\
\Pi&=&\left(\begin{array}{ccc}
  \frac{\pi^0}{F_\pi}+\frac{\eta}{\sqrt3 F_\eta} &-\sqrt2
 \frac{\pi^+}{F_\pi} & -\sqrt2 \frac{K^+}{F_K}\\
  -\sqrt2 \frac{\pi^-}{F_\pi}& -\frac{\pi^0}{F_\pi}+
  \frac{\eta}{\sqrt3 F_\eta} & -\sqrt2 \frac{K^0}{F_K}\\
  -\sqrt2 \frac{K^-}{F_K}&-\sqrt2 \frac{\bar K^0}{F_K} & -
  \frac{2\eta}{\sqrt3 F_\eta}
                 \end{array} \right),
\ear
with decay constants normalized as $F_\pi=93.3$~MeV, $F_K\approx 1.2 F_\pi$.

In the next $O(p^4)$ order the interactions of the (pseudo)Goldstone
mesons are described by the following EChL (we write only terms surviving
in the large $N_c$ limit)
\cite{GLsu3}\footnote{The fourth order EChL without symmetry breaking
term containing derivatives was analyzed for the first time in
ref.~\cite{VVV}}:

  \bear {\cal L}^{(4)}&=& (2L_2+L_3)\tr(L_\mu L^\mu
 L_\nu L^\nu ) + L_2  \tr(L_\mu L_\nu L^\mu L^\nu ) \nonumber \\ &+&
L_5 \tr(L_\mu  L^\mu \chi)+ L_8\tr(\chi^2).  \la{echl4} \ear
For the parameters of the  fourth order  EChL we use
here notations of Gasser and Leutwyler \cite{GLsu3}. We see that
the fourth order EChL has in the large $N_c$ limit four
independent parameters.\footnote{Without taking the large $N_c$
limit it depends on eight parameters \cite{GLsu3}}

All sixth order terms of the EChL were classified in ref. \cite{feasch}.
For  our analysis of the analytical sixth order chiral
contributions to the scattering amplitudes we need terms which
are not vanishing in the leading order of the $1/N_c$ expansion
and contribute to the pion--pion and pion--kaon scattering. This
lagrangian has the form:

\bear
 {\cal L}^{(6)}&=&
K_1 \tr(\partial_\sigma  L_\mu \partial^\sigma L^\mu L_\nu L^\nu) +
K_2 \tr(\partial_\sigma  L_\mu \partial^\sigma L^\nu L^\mu L_\nu)
\nonumber \\
& +& K_3 \tr(\partial_\sigma  L_\mu \partial^\sigma L^\nu L_\nu L^\mu) +
K_4 \tr(\partial_\sigma  L_\mu L_\nu \partial^\sigma L^\mu L^\nu)
\nonumber \\
&+&K_5 ( \tr(\partial_\sigma  L_\mu L^\mu \partial^\sigma L^\nu L_\nu ) +
 \tr(\partial_\sigma  L_\mu L_\nu \partial^\sigma L^\nu L^\mu))
\nonumber \\
& + &K_6 \tr (\chi L_\mu L^\mu L_\nu L^\nu) +
K_7 \tr (\chi L_\mu L_\nu L^\mu L^\nu) +
K_8 \tr (\chi L_\mu L_\nu L^\nu L^\mu)
\nonumber  \\
&+&K_9 \tr (L_\mu L^\mu \chi^2) +K_{10} \tr(L_\mu \chi L^\mu \chi)+
K_{11} \tr(L_\mu \bar\chi L^\mu \bar\chi)
\nonumber  \\
&+& K_{12} \tr(L_\mu  L^\mu \bar\chi \bar\chi)
+K_{13} \tr (\chi \bar \chi^2) +K_{14} \tr (\chi^3),
\label{echl6}
\ear
where $\bar\chi = 2 B_0 (\hat{m}U-U^\dagger\hat{m} )$ and
$\chi$, $\hat{m}$ and $L_\mu$ are defined after eq.(\ref{echl2}).
New coupling constants $K_{1 \ldots 14}$ determine the polynomial
(analytical) part of low-energy behaviour of the two- and
four-point meson Green functions in the large $N_c$ limit (OZI
rule).  Non--analytical  part of the Green
functions and the violation of the OZI rule arise from mesonic
loops. The non--analytic contributions to the $\pi\pi$ and $\pi
K$ scattering amplitudes to one loop were calculated in
refs.~\cite{GL,BerKaiMei}, part  of two loops contributions to
$\pi\pi$ scattering
were calculated recently in refs.~\cite{Col},
the low energy $\pi\pi$ amplitude to one and two loops
were recently obtained in generalized chiral perturbation theory
\cite{KneMouSteFuch},
the complete
two--loop calculations in standard chiral perturbation
theory will be finished soon \cite{two}.  To calculate the
$O(p^6)$ corrections to the low--energy scattering amplitudes
completely one needs (along with two loops contributions) to
know the parameters $K_{1 \ldots 14}$ of the sixth order EChL.
Apart from `` practical" value, the determination of the sixth
order EChL parameters $K_{1 \ldots14}$ has, we think, a wider
theoretical significance since these constants are related to
fine features of the spontaneous chiral symmetry breaking in QCD
and they can be used to check predictions of various models for
chiral symmetry breaking in QCD.

\section{ Low-energy $\pi\pi$ and $\pi K$ scattering amplitudes}
\setcounter{equation}{0}
\label{ampl_sect}

In this section we calculate the polynomial part of the chiral
corrections to masses, decay constants and $\pi\pi$ and $\pi K$ scattering
amplitudes in terms of the parameters of the sixth order EChL
given by eqs.~(\ref{echl2},\ref{echl4},\ref{echl6}). We
follow closely  the technique described in
refs.~\cite{GL,GLsu3,BerKaiMei}, hence we give below only the
results of our calculations without technical details.

\subsection{Masses and decay constants}

Masses and decay constants of pions and kaons are extracted from
two point correlation function of axial currents. The result of
calculations for polynomial part of the chiral corrections is:

\bear
M_\pi^2&=&2 m B_0 \bigl\{1 + m\frac{16 B_0}{F_0^2} (2L_8-L_5) -
m^2 \frac{236 B_0^2}{F_0^4}L_5 (2L_8-L_5)\nonumber \\
&+&m^2 \frac{64 B_0^2}{F_0^2}(K_9+K_{10}+2K_{13}+3K_{14})+O(m^3)\bigr\},\\
M_K^2&=& (m+m_s)B_0\bigl\{1+(m+m_s)\frac{8B_0}{F_0^2}(2L_8-L_5)-
(m+m_s)^2 \frac{64 B_0^2}{F_0^4}L_5 (2L_8-L_5)\nonumber \\
&+&\frac{32 B_0^2}{F_0^2}
( (m^2+m_s^2)(K_9+K_{13}+3K_{14})+m m_s(2K_{10}+2K_{13}))\bigr\},\\
F_\pi^2&=&F_0^2\bigl\{1+m\frac{16 B_0}{F_0^2}L_5 - \frac{64 m^2
B_0^2}{F_0^2} (K_9+K_{10})+O(m^3) \bigr\}, \\
F_K^2&=&F_0^2\bigl\{1+(m+m_s)\frac{8
B_0}{F_0^2}L_5 - \frac{32 B_0^2}{F_0^2} ((m^2+m_s^2)K_9+2m m_s
K_{10})\bigr\},
\ear

\subsection{Scattering amplitudes }

Let us consider the elastic $\pi \pi$--scattering process

$$
\pi_a(k_1)+\pi_b(k_2)\rightarrow\pi_c(k_3)+\pi_d(k_4).
$$
( $a,b,c,d=1,2,3$ are the isotopic indices and
$k_1,..,k_4$ --- pion momenta.)
Its amplitude $ M^{abcd} $  can be written in the form:

\beq
M^{abcd}=\delta^{ab}\delta^{cd}A+\delta^{ac}\delta^{bd}B+
\delta^{ad}\delta^{bc}C ,
\label{A-def}
\eeq
where $A,B,C$ are the scalar functions of Mandelstam variables
 $s,t,u$:

\beq
s=(k_1+k_2)^2,\qquad  t=(k_1-k_3)^2,\qquad  u=(k_1-k_4)^2,
\la{mv} \eeq
obeying the Bose--symmetry requirements:

\begin{eqnarray}
A(s,t,u) &=& A(s,u,t) \ , \nonumber \\
B(s,t,u) &=& A(t,s,u) \ , \label{A-def2} \\
C(s,t,u) &=& A(u,t,s) \ . \nonumber
\end{eqnarray}
The amplitude of the $\pi K$ scattering process

\beq
\pi_a(k_1) +K_\alpha(k_2) \rightarrow \pi_b(k_3) +K_\beta(k_4).
\nonumber
\eeq
can be expressed in terms of two (iso)scalar functions $T_{+}(\nu,t) $
and $T_{-}(\nu,t)$ by

\beq
\nonumber
M^{ab}_{\alpha \beta}=\delta^{ab} \delta_{\alpha \beta} T_{+}(\nu,t)+
i \epsilon^{abc} \sigma^c_{ \beta\alpha}T_{-}(\nu,t),
\eeq
where invariant variable $\nu=s-u$ is expressed via Mandelstam variables
\eq{mv}.
At low momenta one can expand the (iso)scalar
amplitudes,  $T_{+}(\nu,t)$ and $T_{-}(\nu,t)$
in power series of invariant kinemantical variables:

\bear
 A(s,t)&=&\sum_{i,j}^{} A_{ij}(m) s^i t^j , \label{A-exp}\\
T_{+}(\nu,t)&=&\sum_{k,l}t_{2k \; l}^{+}(m_s) \nu^{2k} t^l,\\
T_{-}(\nu,t)&=&\sum_{k,l}t_{2k+1 \; l}^{-}(m_s) \nu^{2k+1} t^l.
\ear
Non-analytical parts of the amplitudes (like $E^4 log(E)$) are
suppressed by additional factors of $1/N_c$. Parameters of the
near threshold expansion  depend on quark masses.  From the
Effective Chiral Lagrangian,
eqs.~(\ref{echl2},\ref{echl4},\ref{echl6}), one gets
an expression for the low-energy parameters of the $\pi \pi $
and $\pi K$ scattering amplitudes as  series in quark mass:\\
{\it \underline{$\pi\pi$ parameters}}\footnote{Corresponding expressions
for the $\pi\pi$ scattering lengths and slope parameters are
given in Appendix~A.}

\bear
\nonumber
A_{00}&=&-\frac{2 m B_0}{F_0^2}
+ \frac{64 m^2 B_0^2}{F_0^4} (3L_2+L_3)\\
\nonumber
&+&\frac{64 m^3 B_0^3}{F_0^4} \biggl\{
2 (K_1+K_2+K_3-K_4-2 K_5) \\
\nonumber
&+& (\tilde{K_6}+\tilde{K_7}+\tilde{K_8} +16 (3 L_2+L_3)(2L_8-L_5))\\
&+&
(4 K_9+4 K_{10}-8K_{11}+8K_{12} +8 K_{13}+6 K_{14}+64 L_5 (3L_5-4L_8))
\biggr\}
\label{a00},\\
A_{01}&=&- \frac{64 m B_0 L_2}{F_0^4}  -\frac{64 m^2 B_0^2 }{F_0^4}
\biggl\{ (K_2+K_3-2 K_5)+ 2 \tilde{K_7}+16 L_2 (2L_8-L_5)
\biggr\},
\label{a01} \\
A_{02}&=&\frac{8 L_2}{F_0^4} +\frac{8 m B_0}{F_0^4}
\biggl\{(K_2+K_3-2 K_5)+ 2 \tilde{K_7}\biggr\},
\label{a02} \\
A_{03}&=&0,
\label{a03} \\
\nonumber
A_{10}&=&\frac{1}{F_0^2} - \frac{32 m B_0}{F_0^4}(2L_2+L_3) \\
\nonumber
&+&\frac{32 m^2 B_0^2}{F_0^4} \biggl\{
  (-3K_1-3K_2-K_3+6K_4+4K_5) \\
&-&2(\tilde{K_6}+\tilde{K_8})-16(2L_2+L_3)(2L_8-L_5)
-2(K_9+K_{10}-4K_{11} +K_{12}) \biggr\},
\label{a10} \\
A_{11}&=& \frac{8L_2}{F_0^4}+\frac{16 m B_0}{F_0^4} \biggl\{
2K_2+K_3-3K_4-2K_5+\tilde{K_7} \biggr\} ,
\label{a11} \\
A_{12}&=&A_{21}=\frac{1}{F_0^4} (-3K_2-K_3+6K_4+2K_5),
\label{a12} \\
A_{20}&=&\frac{4(2L_2+L_3)}{F_0^4}+\frac{8mB_0}{F_0^4 }\biggl\{
(3K_1+3K_2-6K_4-4K_5) +\tilde{K_6}+\tilde{K_8} \biggr\},
\label{a20} \\
A_{30}&=&-\frac{2}{F_0^4}(K_1+K_2-2K_4-2K_5),
\label{a30}
\ear
where we introduce the following notations:

\bear
\nonumber
\tilde{K_6}&=&K_6-\frac{8}{F_0^2}(2L_2+L_3)L_5, \\
\nonumber
\tilde{K_7}&=&K_7-\frac{16}{F_0^2}L_2L_5, \\
\nonumber
\tilde{K_8}&=&K_8-\frac{8}{F_0^2}(2L_2+L_3)L_5.
\ear
The result for the low--energy parameters of the $\pi K$ scattering
amplitude (with $m_u=m_d=0$) is the following:\\
{\it \underline{$\pi K$ parameters}}

\bear
t^{+}_{00}&=&0 \qquad (\mbox{ exactly}) \label{exact1} \\
t^{+}_{01}&=&\frac{1}{4 F_0^2}- \frac{4 m_s B_0}{F_0^4}(2L_2+L_3)\\
&+&\frac{ m_s^2 B_0^2}{F_0^4} \biggl\{-2 K_1-4 \tilde{K_8}+
8 (K_9-K_{12})-64(2L_2+L_3)(2L_8-L_5) \biggr\} \\
t^{+}_{02}&=&\frac{12L_2+5L_3}{2 F_0^4}+ \frac{ m_s B_0}{4F_0^4}
\biggl\{(K_2-3K_3-8K_5)+ 2(\tilde{K_6}+\tilde{K_7}+4\tilde{K_8})
\biggr\} \\
t^{+}_{03}&=&\frac{1}{8 F_0^4}(-7K_1-K_2+2K_3+2K_4+10K_5) \\
t^{+}_{20}&=&\frac{4L_2+L_3}{2 F_0^4}+ \frac{ m_s B_0}{4F_0^4}
\biggl\{(K_2+K_3)+ 2(\tilde{K_6}+\tilde{K_7})
\biggr\} \\
t^{+}_{21}&=&\frac{1}{8 F_0^4}(3K_1-3K_2-2K_3+6K_4-2K_5) \\
t^{-}_{10}&=&\frac{1}{4F_0^2} \qquad \mbox{(exactly!)}\label{exact2}  \\
t^{-}_{11}&=&\frac{-L_3}{ F_0^4}+ \frac{ m_s B_0}{2F_0^4}
\biggl\{(K_2+K_3-4K_5)+ 2(\tilde{K_7}-\tilde{K_6})
\biggr\} \\
t^{-}_{12}&=&3t^{-}_{30}=\frac{3}{8 F_0^4}(-K_1-K_2+2K_3+2K_4).
\label{tm12}
\ear
Let us stress that the parameters of the $\pi K $ scattering amplitude
 $t^+_{00}$ and $t^-_{10}$ given by eqs.(\ref{exact1},\ref{exact2})
have no corrections due to non--zero strange quark mass in any order of
$m_s$ expansion, though there are corrections of order $O(m \cdot m_s)$
and $O(m_s/N_c)$. The former corrections are expected to be very
small, whereas the latter appears due to loop contributions and
are not considered in the present paper since we caclulate
only polynomial part of the chiral corrections to the
scattering amplitudes.
These \underline{exact} on-shell low energy theorems will
enable us to fix parameters of the dual resonance models for
$\pi\pi$ and $\pi K$ scattering amplitudes.

\section{ Dual resonance models for scattering of the  (pseudo) Goldstone
          particles}
\setcounter{equation}{0}
\label{dual_sect}

The dual resonance models were invented in 60's to describe some striking
features of hadron interactions. To good accuracy the mesons and baryons
lie on linear Regge trajectories, a wealth of high energy scattering data
is modelled very well by single reggeon exchange
(for a review see ref.~\cite{Regge_review}).
Later it has been found that the dual resonance models
follow from  string theory and that duality is a consequence
of  the infinite-dimensional conformal symmetry of  string
theories.  After the advent of the Quantum Chromodynamics much
evidences have been found that QCD in large $N_c$ limit might be
equivalent to some string theory.  Let us just list
them:

\begin{itemize}
\item The success of  Regge phenomenology
\cite{Regge_review}
 \item The perturbation expansion in the large $N_c$
limit of  QCD can be written as a sum over surfaces
which may correspond to a sum over string world sheets
\cite{tHooft74}
\item
The strong coupling expansion for lattice gauge theory strongly resembles a
string theory
 \cite{lattice}
\item
The Wilson loop expectation values in the large $N_c$ limit
satisfy equations which is equivalent to those for one or
another specific string theory strings \cite{Migdal_Makeenko}
\item
2D QCD can be rewritten as a string theory \cite{Gross}
\end{itemize}
Unfortunately, the precise form of the string theory
corresponding to QCD is unknown.  In the present paper we
exploit well established facts about  QCD to find conditions
which are imposed on  string theories (dual resonance models)
by these facts. One of the most prominent phenomena occurring in
 QCD is the spontaneous breaking of chiral symmetry. It
leads to numerous  low--energy theorems for scattering
amplitudes of the (pseudo)Goldstone particles ($\pi$, $K$ and
$\eta$), which are written compactly in terms of the  effective
chiral lagrangian, eqs.(\ref{echl2},\ref{echl4},\ref{echl6}).

A dual resonance model for the $\pi\pi$ scattering amplitude
consistent with low-energy theorems in the chiral limit and
ghost free has been suggested in \cite{Lovelace,Shapiro}. It has
a form (in the chiral limit):

\bear
\nonumber
M^{abcd}&=& \tr(\tau^a\tau^b \tau^c \tau^d) V(s,t) + \mbox{non-cyclic
permutations}, \\
V(s,t)&=& \lambda
\frac{\Gamma(1-\alpha_\rho(s))\Gamma(1
-\alpha_\rho(t)))}{\Gamma(1-\alpha_\rho(s)-\alpha_\rho(t))},
\label{Lovelace_amplitude}
\ear
where the $\rho$-meson Regge trajectory and the constant $\lambda$ are
chosen to be
 \bear \nonumber
 \alpha_\rho(s)&=&\frac12 + \frac{s}{2 m_\rho}, \\
\lambda&=&-\frac{m_\rho^2}{\pi F_0^2},
\ear
in order to ensure the low-energy theorem for the amplitude:

\beq
\lim_{s,t \rightarrow 0}A(s,t)=\frac{s}{F_0^2} + O(p^4).
\eeq
The ampltude (\ref{Lovelace_amplitude}), besides correct low--energy
properties, satisfies a Regge asymptotic restrictions at high energies.
Moreover, the positions and residues of the resonance poles are
in a good agreement with phenomenological ones. Hence the basic
phenomenological features of the hadron interactions are
implemented by a simple dual amplitude
(\ref{Lovelace_amplitude}).

Performing low--energy expansion of  the amplitude
(\ref{Lovelace_amplitude}) one can immediately  extract parameters $L_1$,
$L_2$ and $L_3$ of the fourth order EChL eq.(\ref{echl4}):

\bear L_2&=&2L_1, \label{nc} \\
L_3&=&-2 L_2
\label{anomaly} \\
L_2&=& \frac{F_0^2}{8 m_\rho^2}
\mbox{ln}(2)\approx 1.25\times 10^{-3}.
\label{l2value}
\ear
The first relation eq.~(\ref{nc}) is identical to one following
from the large $N_c$ conditions for the meson scattering
amplitude\cite{GLsu3}.  These conditions are ``built in'' in the
dual resonance models through the Chan--Paton isotopic factor.
The second relation, eq.~(\ref{anomaly}), is exactly the
relation predicted by integration of the non-topological chiral
anomaly \cite{DiaEid,And,Bal,DiaPet} and it holds with any type
of satellites added to a simple Lovelace--Shapiro amplitude
(\ref{Lovelace_amplitude}) \cite{BolManPolVer}. Moreover the numericl value
of $L_2$ (\ref{l2value}) is close to that found by Gasser and Leutwyler
in ref.~\cite{GLsu3} $L_2=(1.7 \pm 0.7)\cdot 10^{-3}$, to recent
determination of this constant from analysis of the $K_{l4}$
decay \cite{BijColGas} $L_2=(1.35 \pm 0.3)\cdot 10^{-3} $ and
simultaneously to that obtained by integration of the
non--topological chiral anomaly $L_2=1.58 \cdot 10^{-3}$.
The values of the combination  $2L_2+L_3$ obtained
in refs.~\cite{GLsu3,BijColGas} are consistent with
zero\footnote{For the direct check of the relation $2L_2+L_3=0$
dictated by non-topological chiral anomaly of QCD and dual
(string) models one need to repeat the fitting procedure of
ref.~\cite{BijColGas} using, among others, variable $2L_2+L_3$}.

To estimate other parameters entering eqs.(\ref{echl4},\ref{echl6}) one has
to introduce an explicit chiral symmetry breaking to the dual resonance
model. In a pioneering works of Lovelace \cite{Lovelace} and
Shapiro \cite{Shapiro} it was achieved by shifting the intercept
of the $\rho$-meson Regge trajectory from $\frac12$ to
$\frac12-M_\pi^2/2m_\rho^2$ to reproduce the Adler zero.
Unfortunately, the Adler condition is an {\it off mass shell}
one, whereas the dual (string) amplitudes can be defined and
 constructed consistently only {\it on mass shell }, and a
continuation of those to unphysical region is ambiguous. Here we
shall use a new way of introducing quark masses (explicit chiral
symmetry breaking) into the dual model. Instead of using the
Adler condition  we shall impose {\it on mass shell} low energy
theorems (like those given by eqs.~(\ref{exact1},\ref{exact2}) )
on the dual amplitudes for the $\pi\pi$ and $\pi K$ scatterings.

\subsection{Dual resonance model for the $\pi\pi$ and $\pi K$
 scattering amplitudes. Non-zero quark masses}

Now we generalize the dual
amplitude for pions (\ref{Lovelace_amplitude}) for a case of small non-zero
quark masses, in this case one can write generically:

\bear
\nonumber
V(s,t)&=&-\frac{m_\rho^2}{\pi F_0^2}(1+a_1 m +a_2 m^2 +\ldots)
\biggl\{
\frac{\Gamma(1-\alpha_\rho(s))\Gamma(1
-\alpha_\rho(t))}{\Gamma(1-\alpha_\rho(s)-\alpha_\rho(t))} \\
&+&(b_1 m + b_2 m^2+ \ldots)
\frac{\Gamma(1-\alpha_\rho(s))\Gamma(1
-\alpha_\rho(t))}{\Gamma(2-\alpha_\rho(s)-\alpha_\rho(t))}
\biggr\},
\label{pipi_parameters_free}
\ear
where the intercept of the $\rho$-meson Regge trajectory
has also mass corrections:

\beq
\alpha_\rho(s)=\frac12 (1+ i_1 m+ i_2 m^2+\ldots)+\frac{s}{2 m_\rho^2},
\label{rho_trajectory}
\eeq
here $m_\rho$ is a mass of the $\rho$ meson in the chiral limit
and coefficients $i_k$ describes corrections to the intercept of the
$\rho$ meson trajectory and simultaneously the quark mass corrections to
the $m_\rho$.  We do not include corrections to the slope of the trajectory
because they can be absorbed by a redifinition of the $\rho$
mass.

It is easy to see
that in the chiral limit the amplitude,
eq.~(\ref{pipi_parameters_free}), coincides with
(\ref{Lovelace_amplitude}). For the second term in
eq.~(\ref{pipi_parameters_free}) we choose the simplest possible satellite
term having no poles at zero momenta. The generalization for
arbitrary satellites is straightforward but unnecessary for our
purposes.  Some of the unknown parameters  $a_i$ and $b_i$ can
be fixed by the low-energy theorems, eqs.~(\ref{a00},
\ref{a10}); the resulting amplitude has the form:

\bear
\nonumber
V(s,t)&=&-\frac{m_\rho^2}{\pi F_0^2}(1-\frac{4 m B_0\ln 2}{m_\rho^2}
(1+\frac{(2\alpha_\rho(0)-1) m_\rho^2}{M_\pi^2})
 +a_2 m^2
+\ldots) \\
\nonumber
&\times&\biggl\{
\frac{\Gamma(1-\alpha_\rho(s))\Gamma(1
-\alpha_\rho(t))}{\Gamma(1-\alpha_\rho(s)-\alpha_\rho(t))} \\
&+&(\frac{ 2 m B_0}{m_\rho^2}+(2\alpha_\rho(0)-1) + b_3 m^3+O(m^4))
\frac{\Gamma(1-\alpha_\rho(s))\Gamma(1
-\alpha_\rho(t))}{\Gamma(2-\alpha_\rho(s)-\alpha_\rho(t))}
\biggr\}.
\label{pipi_parameters_fixed}
\ear
The mass corrections to the intercept $\alpha_\rho(0)$ and
parameters $a_2$ and $b_3$ are not fixed by the on--mass--shell
low-energy theorems.  The {\it off mass shell} Adler
conditions being imposed on the amplitude
eq.~(\ref{pipi_parameters_fixed}) give the following relation
\cite{Lovelace,KawKitYab}:

 \beq
(2\alpha_\rho(0)-1)=i_1 m+i_2 m^2+\ldots=-\frac{M_\pi^2}{m_\rho^2}.
\label{Adler_conditions}
\eeq
We shall keep the parameters $i_k$ (or, equivalently,
$\alpha_\rho(0)$) free, because to implement the Adler
conditions one has to know the continuations of the dual
amplitude off mass shell. The latter problem is not solved in
the dual (string) models.

Now let us construct a dual amplitude for the $\pi K$ elastic
scattering for the simplified case of $m_u=m_d=0$ and $m_s \neq
0$. In this case we have very powerful low--energy theorems
(\ref{exact1},\ref{exact2}) which, as we shall see, fix
parameters of the dual amplitude completely.

The dual $\pi K$ scattering amplitude depends on the $\rho$-meson
trajectory (\ref{rho_trajectory}) and $K^*$ one, the  latter
generically has the form (we assume that $m_u=m_d=0$ ):

\beq \alpha_{K^*}(s)=\frac12 (1+ j_1 m_s+
j_2 m_s^2+\ldots)+ \frac{s}{2 m_\rho^2}(1+n_1 m_s+n_2 m_s^2 +\ldots),
\eeq
where we introduce, on general grounds, the quark
mass corrections to the intercept and the slope of the
$K^*$-meson trajectory. The $K^*$ meson mass (in large $N_c$
limit) is determined by the equation:  \beq
\alpha_{K^*}(m_{K^*}^2)=1.
\eeq

The dual resonance amplitude for the $\pi K$ elastic
scattering satisfying all the on mass shell low--energy theorems
has a form:

\bear
\nonumber
T_{\pm}(\nu,t)&=& V_{\rho K^*}(s,t) \pm V_{\rho K^*}(u,t), \\
\nonumber
V_{\rho K^*}(s,t)&=&-\frac{m_\rho^2}{2\sqrt\pi F_0^2}
\frac{\Gamma(1+
\frac{m_{K^*}^2-m_\rho^2-M_K^2}{2 m_\rho^2} ) }{\Gamma(\frac12+
\frac{m_{K^*}^2-m_\rho^2-M_K^2}{2 m_\rho^2} ) }
\biggl\{ \frac{\Gamma(1-\alpha_{K^*}(s))\Gamma(1
-\alpha_\rho(t))}{\Gamma(1-\alpha_{K^*}(s)-\alpha_\rho(t))} \\
&-&\frac{m_{K^*}^2-m_\rho^2-M_K^2}{2 m_\rho^2}
\frac{\Gamma(1-\alpha_{K^*}(s))\Gamma(1
-\alpha_\rho(t))}{\Gamma(2-\alpha_{K^*}(s)-\alpha_\rho(t))}
\biggr\},
\label{venpik}
\ear
where we take into account the fact that the low--energy
theorems can be satisfied only with the following relations between
parameters of the $\pi\pi$ and $\pi K$ dual amplitudes:

\bear
\nonumber
i_1&=&2 j_1, \\
n_k&=&0 \qquad \mbox{universality of the Regge trajectories slope !}.
\ear
It is remarkable that soft pion theorems lead to the universality of the
Regge trajectories slopes and also give a mass relation

\beq
\frac{m_{K^*}^2-m_\rho^2}{ m_\rho^2}
=-(2 \alpha_\rho(0)-1) \frac{m_s}{2 m}
\approx - \frac{i_1 m_s}{2}.
\label{mass_formula}
\eeq
{}From this mass formula and Adler relation
eq.~(\ref{Adler_conditions}) one gets the
famous mass relation of Lovelace \cite{Lovelace} :
\beq
m_{K^*}^2=m_\rho^2 + M_\pi^2.
\eeq

By virtue of eq.~(\ref{mass_formula}) we shall use the mass difference
$M_{K^*}^2-m_\rho^2$ as a free parameter equivalent to
$\alpha_\rho(0)$.

\section{Low--energy expansion of the dual resonance amplitude}
\setcounter{equation}{0}
\label{low_dual_sect}

Expanding the dual resonance amplitudes,
eqs.~(\ref{pipi_parameters_fixed}, \ref{venpik}), for $\pi \pi$
and $\pi K$ scatterings at low momenta one can obtain
some of the EChL parameters.
In order to fix the others one needs to know additionally dual
 n-point amplitudes for pions.  Unfortunately, less is known
 about n-point generalization of the Lovelace formula
(\ref{Lovelace_amplitude})\footnote{ See, though, a recent paper
\cite{Slava} where the n-point generalization of the
Lovelace--Shapiro amplitude were suggested}.  Also the dual
amplitudes in question depend on parameter(s) $\alpha_\rho(0)$
(or, equivalently, $i_k$) which are not fixed by soft pion
theorems.  We choose $\alpha_\rho(0)$ corresponding to the
experimental values of the vector meson masses in
eq.~(\ref{mass_formula}).

Expanding the dual resonance amplitudes,
eqs.(\ref{pipi_parameters_fixed}, \ref{venpik}), at low energies
and comparing the result with eqs.~(\ref{a00}-\ref{a30}) and
eqs.~(\ref{exact1}-\ref{tm12}) respectively one can fix
the following parameters of the sixth order EChL:

\bear
     \nonumber
K_1&=&\quad \! 0, \\
     \nonumber
K_2&=&
-\frac{F_0^2( \pi^2+15 \ln^22)}{60 m_\rho^4},  \\
     \nonumber
K_3&=& \quad \!
\frac{F_0^2 \pi^2}{80 m_\rho^4},  \\
     \nonumber
K_4&=&
-\frac{F_0^2 \pi^2}{96 m_\rho^4},  \\
     \nonumber
K_5&=& \quad \!
\frac{F_0^2 \pi^2}{80 m_\rho^4},  \\
\label{six_order_parameters}
\tilde{K_6}&=&
-\frac{F_0^2 \pi^2}{80 m_\rho^4}-\frac{F_0^2 \pi^2}{64 m_\rho^4}
\frac{m_{K^*}^2-m_\rho^2-M_K^2}{M_K^2},  \\
     \nonumber
\tilde{K_7}&=&\quad \!
\frac{F_0^2 (7\pi^2+60 \ln^2 2)}{480 m_\rho^4}-\frac{F_0^2 \pi^2}{192
m_\rho^4} \frac{m_{K^*}^2-m_\rho^2-M_K^2}{M_K^2},  \\
     \nonumber
\tilde{K_8}&=&\quad \!
\frac{3 F_0^2 \pi^2}{160 m_\rho^4}+\frac{F_0^2 \pi^2}{64 m_\rho^4}
\frac{m_{K^*}^2-m_\rho^2-M_K^2}{M_K^2}, \\
     \nonumber
K_9-K_{12}&=& \quad \!
\frac{3 F_0^2 \pi^2}{320 m_\rho^4}+\frac{F_0^2 \pi^2}{128 m_\rho^4}
\frac{m_{K^*}^2-m_\rho^2-M_K^2}{M_K^2}, \\
     \nonumber
K_{10}-K_9&=&-\frac{F_0^2 \pi^2}{128 m_\rho^4}
-\frac{F_0^2 \pi^2}{128 m_\rho^4}
\frac{m_{K^*}^2-m_\rho^2-M_K^2 }{M_K^2},
\ear
This is the main result of the paper.
 Numerically from  these
equations one has (taking $m_{K^*}=892$~MeV):

\bear
\nonumber
 K_1
m_\rho^2&=& \quad \! 0,\\
\nonumber
 K_2m_\rho^2&\approx& -3.72\cdot10^{-3} ,   \\
\nonumber
K_3m_\rho^2&\approx&\quad \! 1.61 \cdot10^{-3},   \\
 \nonumber
K_4m_\rho^2&\approx& -1.34\cdot10^{-3},   \\
\label{anomaly6_numeric}
K_5m_\rho^2&\approx&\quad \! 1.61 \cdot10^{-3},   \\
\nonumber
\tilde{K_6}m_\rho^2 &\approx& -1.26 \cdot10^{-3},  \\
\nonumber
\tilde{K_7}m_\rho^2 &\approx&\quad \! 2.78 \cdot10^{-3},  \\
\nonumber
\tilde{K_8}m_\rho^2 &\approx&\quad \! 2.07 \cdot10^{-3},  \\
     \nonumber
(K_9-K_{12}) m_\rho^2 &\approx&\quad \! 1.03 \cdot10^{-3}, \\
     \nonumber
(K_{10}-K_9)m_\rho^2 &\approx& -0.82 \cdot10^{-3},
  \ear
In these numerical estimates for the parameters $m B_0$,
$F_0$ and $2L_8-L_5$ we use the values given by Gasser and
Leutwyler \cite{GL,GLsu3}:

\bear
F_0&=&88 \mbox{MeV}, \\
2 m B_0&=& (141)^2 \mbox{MeV}^2,\\
m_s B_0&=& (505)^2 \mbox{MeV}^2,\\
2L_8-L_5&=&(0 \pm 1.1) 10^{-3}.
\label{num_masses}
\ear

In order to fix other coefficients  we have to analyze not only
scattering amplitudes but also mass and decay constants
splittings. This can be done consistently only if two
loop contributions to these quantities are taken into account,
so this deserves further study. To estimate  the
polynomial contributions of the sixth order to $S-$ and $P-$wave
$\pi \pi$ scattering lengths one needs to pin down  the following
combination of the parameters $-K_9-K_{10}+4K_{11}-2K_{12}$,
this is equivalent to fixing parameters $a_2$ and $b_3$ in the
dual resonance $\pi \pi $ scattering  amplitude
(\ref{pipi_parameters_fixed}).
{}From  eqs.~(\ref{anomaly6_numeric}) one can assume
tentatively that $|K_{9 \div 14}|\sim  10^{-3}/m_\rho^2$.
We shall use this tentative numbers to
estimate  uncertainties due to unknown parameters.

\subsection{Comparison with Chiral Quark Model}

In the previous section we showed that the low--energy constants
of the {\it fourth} order EChL obtained from the dual resonance
models in the chiral limit coincide with corresponding constants
obtained by integration of the non--topological chiral anomaly
\cite{DiaEid,And,Bal} and from the gradient expansion
of the fermion determinant in the effective chiral quark
model \cite{DiaPet}. Let us compare the low--energy constants
(\ref{anomaly6_numeric} ) with the  corresponding constants
obtained by gradient expansion of the fermion determinant in
the effective chiral quark model.

According to Manohar and Georgi \cite{ManGeo} we can describe
the strong interactions at energies below the scale of chiral
symmetry breaking by a set of fields consisting of $SU(N_f)_V$
multiplet of quarks with a dynamical mass $M$ and Goldstone
bosons. This picture of the low--energy QCD emerges naturally in
the low--momenta limit from the instanton picture of QCD.
According to ref.~\cite{DiaPet} the contents of QCD at
low--momenta comes to dynamically massive quarks interacting
with pseudoscalar fields whose kinetic energy appears only
 dynamically through quark loops.  The basic quantities of the
model, viz. the momentum-dependent quark mass $M(p)$ and the
intrinsic ultra-violet cut-off have been also estimated in
ref.~\cite{DiaPet} through the $\Lambda_{QCD}$ parameter.

The low-momenta QCD partition function is given by the
functional integral over pseudoscalar and quark fields (in the
chiral limit):

\begin{eqnarray}
{\cal Z} &=& \int {\cal D}\Psi {\cal D}\bar \Psi
{\cal  D}\pi^A\:
\exp \left( i\int d^4x  \bar \Psi iD\Psi \right) \\
&=&\int {\cal D} \pi^A\:exp\left( iS_{eff}[\pi]\right ),
\label{Partfunc}
\end{eqnarray}
\begin{equation}
S_{eff}[\pi] \;=\; - \mbox{Sp}\log iD,
\label{EffAct}
\end{equation}
where $iD$ denotes the Dirac differential operator entering the
effective fermion action:

 \bear
 S_{eff}^{ferm}&=& \int d^4x \bar \Psi iD\Psi,
 \label{FermAct} \\
   iD\;&=&\; (- i
\rlap{/}{\partial} + MU^{\gamma_5} ),
\label{Diracop}
\ear
with the pseudoscalar chiral field
\begin{equation}
U^{\gamma_5}=e^{i\pi^A\lambda^A\gamma_5}.
\end{equation}
$\lambda^A$ are Gell-Mann matrices
and $M$ is the dynamical quark mass which arises as a result of the
spontaneous chiral symmetry breaking and is momentum-dependent. The
momentum dependence of $M$ introduces the natural ultra-violet cut-off
for the theory given by eq. (\ref{Partfunc} ).
Performing the expansion of the effective action for pions,
given by the fermion determinant (\ref{EffAct}), in powers of
pion momenta one reveals  the EChL for pions in the large $N_c$
limit.  For the fourth order EChL this gives \cite{DiaPet}:

\bear
\nonumber
L_2 &=& \quad \!\frac{1}{12}\frac{N_c}{24 \pi^2} \approx 1.58\cdot
10^{-3}, \\
\nonumber
L_3&=&-2 L_2.
\ear

 In refs.~\cite{Zuk,PraVal} the low--energy
constants of the sixth order EChL were calculated as functions
of parameters of the effective chiral quark model, i.e. the
 constituent quark mass $M$ and the effective cut--off
 $\Lambda$  (proper--time
 regularization scheme were used):

\bear
\nonumber
K_1&=&
-\frac{3}{10}
 \frac{N_c}{96 \pi^2 M^2}\Gamma(3,\frac{M^2}{\Lambda^2}), \\
\nonumber
K_2&=& \quad \!
\frac{3}{10}
 \frac{N_c}{96 \pi^2 M^2}\Gamma(3,\frac{M^2}{\Lambda^2}), \\
\nonumber
K_3&=& \quad \!\frac{N_c}{96 \pi^2 M^2}
\bigl[\frac{1}{5}\Gamma(2,\frac{M^2}{\Lambda^2})+
\frac{3}{40}\Gamma(3,\frac{M^2}{\Lambda^2}) \bigr], \\
\nonumber
K_4&=& \quad \!\frac{1}{5} \frac{N_c}{96 \pi^2
M^2}\Gamma(3,\frac{M^2}{\Lambda^2}),
\\
\nonumber
K_5&=& \quad \!\frac{N_c}{96 \pi^2 M^2}.
\bigl[\frac{1}{10}\Gamma(2,\frac{M^2}{\Lambda^2})-
\frac{3}{80}\Gamma(3,\frac{M^2}{\Lambda^2}) \bigr],
\ear
where $ \Gamma(n,x)$ is the incomplete gamma function.
We see that it is impossible to reproduce our
values for the corresponding low--energy constants
(\ref{anomaly6_numeric} ), adjusting two parameters of the
effective chiral quark model .  In our view this discrepancy
might be due to neglection of nonlocality of the
corresponding effective fermion action  (\ref{FermAct}) (say,
owing to momentum dependence of the constituent quark mass,
which is predicted, for example, by instanton models of the QCD
vacuum \cite{DiaPet}).  The fourth order EChL parameters are
less sensitive to the nonlocality, while the higher order
ones are strongly dependent on this. Knowing the sixth order
EChL parameters one can find, in principle, the corresponding
effective non-local fermion action of the effective chiral quark
model.

\subsection{ Polynomial contributions to the $\pi\pi$ and
$\pi K$ low--energy scattering paprameters}

Now one can estimate polynomial contributions to the $\pi \pi$
and $\pi K$ scattering parameters due to the sixth order EChL.
To this end one uses values of the parameters given by
eq.~(\ref{six_order_parameters}) and formulae
(\ref{a00app}--\ref{b22app}) from Appendix~A.  The result is the
following:\\
{\it \underline{$\pi \pi$ scattering lengths and slope parameters}}

\bear \label{a00n} a_0^0&=&\quad \! \frac{7 m B_0}{16\pi F_0^2}
\bigl\{1+\frac{4m B_0}{7 m_\rho^2} (5 \ln 2+48 \frac{m_\rho^2}{F_0^2}
 (2L_8-L_5))+
\frac{4m^2 B_0^2\pi^2 h_1}{7 m_\rho^4} +O(m^4)\bigr\}, \\
a_0^2&=& -\frac{ m B_0}{8\pi F_0^2}
\bigl\{1-\frac{4m B_0 \ln 2}{ m_\rho^2} +
\frac{m^2 B_0^2\pi^2 h_2}{ m_\rho^4} +O(m^4)\bigr\}, \\
a_1^1&=& \quad \!\frac{ 1}{24 \pi F_0^2}
\bigl\{1+\frac{8 m B_0 \ln 2}{ m_\rho^2} +
\frac{m^2 B_0^2\pi^2 h_3}{ m_\rho^4} +O(m^4)\bigr\}, \\
a_2^0&=&\quad \! \frac{ \ln 2}{20 \pi F_0^2 m_\rho^2}
+\frac{ m B_0 }{240 \pi F_0^2  m_\rho^4}
(-\pi^2+72 \ln^2 2-\pi^2\frac{m_{K^*}^2-m_\rho^2}{M_K^2})
+O(m^2), \\
a_2^2&=& -\frac{ m B_0 \pi}{120  F_0^2  m_\rho^4}+O(m^2), \\
a_3^1&=&\quad \! \frac{ \pi^2+12 \ln^2 2}{840 \pi  F_0^2  m_\rho^4}+O(m), \\
b_0^0&=&\quad \! \frac{1}{4\pi F_0^2}
\bigl\{1+\frac{4m B_0\ln 2}{ m_\rho^2}+
\frac{m^2 B_0^2\pi^2 h_4}{ m_\rho^4} +O(m^3)\bigr\}, \\
b_0^2&=& -\frac{1}{8\pi F_0^2}
\bigl\{1-\frac{8m B_0\ln 2}{ m_\rho^2}+
\frac{m^2 B_0^2\pi^2 h_5}{ m_\rho^4} +O(m^3)\bigr\}, \\
b_1^1&=&\quad \! \frac{ \ln 2}{6 \pi F_0^2 m_\rho^2}
+\frac{ m B_0 }{18 \pi F_0^2  m_\rho^4}
 (\pi^2+18 \ln^2 2-\pi^2\frac{m_{K^*}^2-m_\rho^2}{4M_K^2})
+O(m^2), \\
b_2^0&=&-\frac{ \pi}{120  F_0^2  m_\rho^4}+O(m), \\
\label{b22n}
b_2^2&=&\quad \! \frac{ -\pi^2+18 \ln^2 2}{120 \pi  F_0^2  m_\rho^4}+O(m),
\ear
where $h_i$ are numbers  not fixed by dual resonance models,
they can be extracted from the analysis of masses and decay
coupling constants of the pseudoscalar mesons.  The constants
$h_i$ are not independent, they are related to each other by the
following relations:

\bear
\label{h_numerical1}
h_3-h_4&=&-\frac{17}{3} -\frac{m_{K^*}^2-m_\rho^2}{6M_K^2} \approx -5.8 , \\
h_4-h_5&=&\quad \, 7 -\frac{m_{K^*}^2-m_\rho^2}{M_K^2}\approx
6.2 , \label{h_numerical2}
\ear
{}From these equations we see that
numerical values of $h_i$ can be quite large ($\sim 10$), hence
we shall use a value  $h_i\approx \pm 10$ to estimate  the
contributions of the sixth order EChL to the $\pi\pi$ scattering
parameters.

One can  also make an order-of-magnitude estimate of the
constants $h_i$ by ``natural'' extension of the dual $\pi\pi$
 amplitude, eq.~(\ref{pipi_parameters_fixed}), with
 substitutions:

\beq
(1-\frac{2 m B_0\ln 2}{m_\rho^2}
(1+\frac{(2\alpha_\rho(0)-1) m_\rho^2}{M_\pi^2})
 +a_2 m^2+\ldots) \rightarrow
\frac{\sqrt\pi
\Gamma(1+\frac{M_\pi^2}{m_\rho^2}x)}{\Gamma(\frac12
+\frac{M_\pi^2}{m_\rho^2} x)} ,
\label{educ_guess}
\eeq
where
$x= \frac{m_{K^*}^2-m_\rho^2-M_K^2}{M_K^2} $.
By this substitution we fix parameters $a_2$ and $b_3$ and
hence $h_i$; the corresponding results are summarized in
Table~I.  We see that $h_i$ can be rather large which is
in agreement with eq.~(\ref{h_numerical1},\ref{h_numerical2}).
It is worth noting that this extension of the dual $\pi\pi$
amplitude, eq.~(\ref{pipi_parameters_fixed}),  (generally
speaking arbitrary) can give us only order-of-magnitude estimate
of the low--energy constants but one can use this estimate   in
qualitative considerations.

Now one can estimate numerically the polynomial contributions to the
$\pi\pi$ scattering parameters, eqs.~({\ref{a00n}-\ref{b22n}),
arising from the sixth order EChL, the corresponding numbers are given in
Table~II (in units of $M_{\pi^+}$). In the same table we
give an experimental values of the scattering parameters taken
 from ref.~\cite{Nag}, though the comparison with an experiment
 is not informative  before loop correction (non-analytical part
of the chiral corrections) added to these quantities.  From
these numerical estimate we see that contributions arising from
the sixth order EChL could be, in principle, as large as the
fourth order ones due to the possibly large values $h_i$.  Using
eqs.~(\ref{h_numerical1},\ref{h_numerical2}) one can calculate
the following combinations of the scattering length and slope
parameters:

 \bear
6 a_1^1-b_0^0&=&\frac{1}{4 \pi
F_0^2} \biggl\{\frac{4mB_0 \ln2}{m_\rho^2}+ \frac{m^2B_0^2
\pi^2}{m_\rho^4}(h_3-h_4) \biggr\} \approx 0.201\cdot
\biggl(0.0458-0.014\biggr),  \\ 2 b_2^0+b_0^0&=&\frac{1}{4 \pi
F_0^2} \biggl\{\frac{12mB_0 \ln2}{m_\rho^2}+ \frac{m^2B_0^2
\pi^2}{m_\rho^4}(h_4-h_5) \biggr\} \approx 0.201\cdot
\biggl(0.137+0.022\biggr), \ear
and indeed we see that the sixth
order contributions could be as large as $10\% \div 25\%$ of the
fourth order ones.

The result for the low--energy parameters of the $\pi K$ scattering
amplitude (with $m_u=m_d=0$) extracted from the dual resonance model
eq.~(\ref{venpik}) is the following:\\
{\it \underline{$\pi K$ parameters}}

\bear
\nonumber
t^{+}_{00}&=&0 \qquad (\mbox{ exactly}) \label{dexact1} \\
\nonumber
t^{+}_{01}&=&\frac{1}{4 F_0^2}+O(m_s^3)\\
\nonumber
t^{+}_{02}&=&\frac{\ln 2}{8 m_\rho^2 F_0^2}+
      \frac{ m_s B_0 \pi^2}{48 m_\rho^4 F_0^2}\cdot
     \frac{m_{K^*}^2-m_\rho^2- M_K^2}{M_K^2}+O(m_s^2), \\
\nonumber
t^{+}_{03}&=&\frac{7 \pi^2+12 \ln^2 2}{384 F_0^2 m_\rho^4}+O(m_s), \\
\label{t30_ven}
t^{+}_{20}&=&\frac{\ln 2}{8 m_\rho^2 F_0^2}
     - \frac{ m_s B_0 \pi^2}{96 m_\rho^4 F_0^2}\cdot
     \frac{m_{K^*}^2-m_\rho^2-  M_K^2}{M_K^2}+O(m_s^2), \\
\nonumber
t^{+}_{21}&=&\frac{- \pi^2+12 \ln^2 2}{128 F_0^2 m_\rho^4}+O(m_s),  \\
\nonumber
t^{-}_{10}&=&\frac{1}{4F_0^2} \qquad \mbox{(exactly!)}\label{dexact2}  \\
\nonumber
t^{-}_{11}&=&\frac{\ln 2}{4 m_\rho^2 F_0^2}
     + \frac{ m_s B_0 \pi^2}{96 m_\rho^4 F_0^2}\cdot
     \frac{m_{K^*}^2-m_\rho^2-M_K^2}{M_K^2}+O(m_s^2), \\
\nonumber
t^{-}_{12}&=&\frac{ \pi^2+12 \ln^2 2}{128 F_0^2 m_\rho^4}+O(m_s),  \\
\nonumber
t^{-}_{30}&=&\frac{ \pi^2+12 \ln^2 2}{384 F_0^2 m_\rho^4}+O(m_s).
     \ear
{}From these expressions we see that in the general dual (string)
model compatible with soft--pions theorem the explicit symmetry
breaking parameter is not $ M_K^2/m_\rho^2$ but rather
$\frac{m_{K^*}^2-m_\rho^2-M_K^2}{m_\rho^2}$. The latter
parameter being of order $\sim m_s \sim M_K^2/m_\rho^2$ has
an additional numerical suppression.

For the $\pi K$ scattering parameters the contributions of the sixth order
     EChL are fixed unambiguously by {\it exact} low--energy
     theorems eqs.~(\ref{exact1},\ref{exact2}).
Substituting numerical values of the parameters eq.~(\ref{num_masses}) into
the eqs.~(\ref{t30_ven}) one gets results showed in
Table~III (in units of $M_{\pi^+}$). In this table we also
show, for completeness, the experimental values of the
  low--energy parameters obtained by Lang and Porod
  \cite{LanPor}. Let us stress again that to compare chiral results with
  experimental data one needs to add to the tree--level results (showed in
  Table~III) the loop corrections.

Surprisingly, the contributions of polynomial part of the sixth
order corrections are rather small (less than 10$\%$ ). Moreover
these corrections are exactly zero if one impose the Adler
conditions eq.~(\ref{Adler_conditions}). The smallness of the
polynomial part of the sixth order contribution to the $\pi
K$  amplitude has been discussed in
ref.~\cite{hep-echl6}.  It has been shown on general grounds
that the low--energy theorems for the $\pi K$ scattering
 are technically respected through the cancellation of
different resonances contributions to $\pi K$ scattering at
low-energies.  Say, for the parameter $t^{+}_{20}$ this
cancellations is not exact (like for $t^{+}_{00}$ and
$t^{-}_{10}$) but nevertheless, even being partial it leads to a
relative smallness of the strange quark mass corrections to this
parameter \cite{hep-echl6}.

\section{Summary and Conclusions}
\setcounter{equation}{0}
\label{concl_sect}

To summarize, we have calculated parameters of the sixth order
effective chiral lagrangian in the large $N_c$ limit from the
dual resonance (string) model for the
scattering amplitudes of the (pseudo)Goldstone particles.
The results are summarized in Table~I.
These parameters determine the polynomial terms in
the low--energy expansion of the  scattering amplitudes up to
the order $O(p^6)$. The polynomial contributions being combined
with  non--analytical parts of the amplitudes arising
from meson loops would enable us to make a precise
calculation of the sixth order contributions to the low--energy
scattering parameters.  From our analysis of the polynomial part
of the sixth order corrections one can conclude that those
corrections to the $\pi\pi$ scattering parameters can be as
large as $10\div 25 \%$ of the fourth order ones, whereas the
analogous corrections to $\pi K$ low-energy scattering
parameters are surprisingly small (usually less than  $10 \%$,
in spite of naive expectation of $M_K^2/m_\rho^2\sim 40 \% $).
The  smallness is explained by ``accidentally'' small
parameter $\frac{m_{K^*}^2-m_\rho^2-M_K^2}{m_\rho^2} \sim 7 \%$
which plays a role of explicit chiral symmetry breaking
parameter in the dual resonance (string) models.

Apart from the ``practical" value, our studies may have wider
theoretical significance. We found that a dual resonance model
with ``built in" soft--pion theorems is consistent with the
non--topological chiral anomaly of the QCD what might be an
indication of the deep relations between QCD and some string
theory.  On other side, application of the soft--pion theorems
to dual resonance models leads to the prediction of the
universality of the  $\rho$- and $K^*$- Regge trajectories
slopes.  Comparing the predictions for the sixth order EChL in
the effective chiral quark model  with ours we
found that the sixth order EChL  from the dual resonance
model differs from that obtained  by gradient expansion of the
fermion determinant in the effective chiral--quark
model\footnote{The corresponding expansion of the fermion
determinant to the fourth order reproduces the non-topological
chiral anomaly results \cite{DiaPet} and so the fourth order of
the gradient expansion is consistent with dual models.}.  In our
view the reason for this difference is that doing gradient
expansion of the fermion determinant \cite{Zuk,PraVal} to the
sixth order one has to take into account a non-locality of the
 effective fermion action (say, the momentum dependence of the
constituent quark mass).  Knowing the sixth order EChL
parameters one can  find, in principle, the corresponding
effective non-local fermion action of the effective chiral quark
model. This work is in a progress.

\section{Acknolegements}

We thank D.~Diakonov for valuable comments. One of us (M.P.) is
grateful to J.~Gasser and H.~Leutwyler for fruitful discussions
and encouraging  to look into this problem. We also profited
from discussions and collaboration on initial stage of this
work with A.~Bolokhov, A.~Manashov and A.~Pron'ko.

\appendix
\section{}
Projecting out amplitudes of definite isospin in $s$-channel
yields:

\bear
M^0(s,t)&=& 3 A(s,t,u)+A(t,u,s)+A(u,s,t), \\
M^1(s,t)&=& A(t,u,s)-A(u,s,t), \\
M^2(s,t)&=& A(t,u,s)+A(u,s,t),
\ear
where $A(s,t,u)$ is defined by eqs.~(\ref{A-def},\ref{A-def2}).
    In the center of mass frame:
\bear
   s&=&4(q^2+M_\pi^2),\nonumber\\
   t&=&-2q^2(1-\cos\theta),\nonumber\\
   u&=&-2q^2(1+\cos\theta),
\label{eq:kin}
\ear
where $q$ is the spatial momentum and $\theta$ is the
scattering angle. We then define the partial wave isospin
amplitudes  according to the following formula:

\bear
M^I(s,t)&=&32\pi
\sum_{l=0}^{\infty}(2l+1)P_l(\cos{\theta})M_l^I (s)\nonumber
\ear

The behaviour of the partial waves  near threshold is of the form
\beq
\mbox{Re}\;M_l^I(s)=q^{2l}\{a_l^I +q^2 b_l^I +O(q^4)\}
\eeq
The quantities $a_l^I$ are referred to as the $\pi\pi$ scattering lengths
and $b_l^I$ as slope parameters.
They can be expressed in terms of the low--energy subthreshold
expansion parameters $A_{kl}$ defined by eq.~(\ref{A-exp}) as
follows:

\bear
a_0^0&=& \frac{1}{32\pi} \bigl\{
5 A_{00}+12 M_\pi^2 A_{10} +48 M_\pi^4 A_{20} +192 M_\pi^6 A_{30}+
O(M_\pi^8) \bigr\}, \label{a00app}\\
a_0^2&=& \frac{1}{16\pi}  A_{00}+O(M_\pi^8) , \\
a_1^1&=& \frac{1}{24\pi} \bigl\{
 A_{10}+4 M_\pi^2 A_{11} +16 M_\pi^4 A_{12} +
O(M_\pi^6) \bigr\}, \\
a_2^0&=& \frac{1}{60\pi} \bigl\{
3 A_{11}+2 A_{20} +32 M_\pi^2 A_{12} +
O(M_\pi^4) \bigr\}, \\
a_2^2&=& \frac{1}{30\pi} \bigl\{
 A_{20} +4 M_\pi^2 A_{12} +
O(M_\pi^4) \bigr\}, \\
a_3^1&=& \frac{1}{35\pi} A_{30} +
O(M_\pi^2) , \\
b_0^0&=& \frac{1}{4\pi} \bigl\{
 A_{10}+2 M_\pi^2 (A_{11}+6A_{20}) +8 M_\pi^4 (A_{12} +9 A_{30})+
O(M_\pi^6) \bigr\}, \\
b_0^2&=& -\frac{1}{8\pi} \bigl\{
 A_{10}-2 M_\pi^2 A_{11}-16 M_\pi^4 A_{12} +
O(M_\pi^6) \bigr\}, \\
b_1^1&=& \frac{1}{6\pi} \bigl\{
 A_{11}-A_{20}+4 M_\pi^2 A_{12} +
O(M_\pi^4) \bigr\}, \\
b_2^0&=& \frac{1}{15\pi} \bigl\{
 5A_{12}-3A_{30} +
O(M_\pi^2) \bigr\}, \\
b_2^2&=& \frac{1}{15\pi} \bigl\{
 2A_{12}-3A_{30} +
O(M_\pi^2) \bigr\}, \label{b22app}
\ear
where we take into account the Bose symmetry requirements :

\bear
A_{21}&=&A_{12}, \\
A_{01}&=&-4 M_\pi^2 A_{02}, \\
A_{02}&=&A_{11}+4 M_\pi^2 A_{21}.
\ear


\begin{table}
\begin{tabular}{|c|c|c|c|}
\hline
& DRM with $M_{K^*}=872 $ MeV & DRM with Adler conditions &
`` extended '' DRM \\
\hline
$K_1$&  $\quad \!0$           & $\quad \!0$ & $\quad \!0 $      \\
\hline
$K_2$ &  $-3.72$      & $-3.72$ &$-3.72$ \\
\hline
$K_3$ &  $\quad \!1.61$       & $\quad \! 1.61$ &$\quad \! 1.61$ \\
\hline
$K_4$ &  $-1.34$       & $-1.34$  &$-1.34$  \\
\hline
$K_5$ & $\quad \! 1.61$         & $\quad \! 1.61$ &  $\quad \! 1.61$ \\
\hline
$\tilde{K}_6 $       & $-1.26$  & $-1.61$ &  $-1.26$     \\
\hline
$\tilde{K}_7$    & $\quad \! 2.78$  & $\quad \!  2.66$ & $\quad \! 2.78$  \\
\hline
$\tilde{K}_8$    & $\quad \! 2.07$ & $\quad \! 2.42$ & $\quad \! 2.07$  \\
\hline
$K_9-K_{12}$ & $\quad \! 1.03$ &$\quad \! 1.21$  &$\quad \! 1.03$ \\
\hline
$K_{10}-K_{12}$  &$\quad \! 0.21$ &$\quad \! 0.21$  &$\quad \! 0.21$ \\
\hline
$K_{11}-K_9$  &-- & --  &    $-2.74$ \\
\hline
$h_1$         &-- & -- &      $\quad \! 4.02$ \\
\hline
$h_2$         &-- & -- &      $\quad \! 0.48$\\
\hline
$h_3$          &-- & -- &      $\quad \! 1.77$\\
\hline
$h_4$          &-- & -- &      $\quad \! 7.70 $\\
\hline
$h_5$          &-- & -- &      $\quad \! 1.54$ \\
\hline
\end{tabular}
\label{t1}
\caption{Low energy coupling constants of the sixth order EChL
eq.~(2.4) $K_i$ in units of $10^{-3}/m_\rho^2$ for different types
of the dual resonance models (DRM) }
\end{table}

\begin{table}
\begin{tabular}{|r|c|c|c|c|c|}
\hline
    & ${\cal L}^{(2)}$&  ${\cal L}^{(4)}$
& ${\cal L}^{(6)}$ &  ${\cal L}^{(6)}$ from`` extended" DRM &
 experiment \cite{Nag}\\
\hline
$a_0^0$&  0.18  &$ (5.9 \pm 6.7)\cdot 10^{-3}$ & $0.27\cdot 10^{-3} h_1$&
$1.09 \cdot 10^{-3} $  & $0.26\pm 0.05$ \\
\hline
$-10 \cdot a_0^2$&  0.51  &$ -2.3 \cdot 10^{-2}$ & $0.14\cdot
10^{-2} h_2$& $0.067 \cdot 10^{-2} $  & $0.28\pm 0.12$  \\
\hline
$10 \cdot a_1^1$&  0.33  &$ 3.1\cdot 10^{-2}$ & $0.09\cdot
10^{-2} h_3$& $0.16 \cdot 10^{-2} $  & $0.38\pm 0.02$ \\
\hline
$10^3 \cdot a_2^0$&  0  & 0.92 & 0.03&
0.03  & $1.7\pm 0.3$ \\
\hline
$10^3 \cdot a_2^2$&  0  & 0 & -0.037 &
-0.037   & $0.13\pm 0.3$ \\
\hline
$10^3 \cdot a_3^1$&  0  & 0 & 0.016 &
0.016  & $0.06\pm 0.02$ \\
\hline
$b_0^0$&  0.20  &$  0.93 \cdot 10^{-2}$ & $0.06\cdot 10^{-2} h_4$&
$ 0.46 \cdot 10^{-2} $  & $0.25\pm 0.03$ \\
\hline
$-10 \cdot b_0^2$&  1.0  &$  -9.3 \cdot 10^{-2}$ & $0.28\cdot 10^{-2}
h_5$& $ 0.43 \cdot 10^{-2} $  & $0.82\pm 0.08$ \\
\hline
$10^2 \cdot b_1^1$&  0  &0.31 & 0.040
&  0.040  & - \\
\hline
$10^5 \cdot b_2^0$&  0  &0 & 7.2 & 7.2   & - \\
\hline
$10^5 \cdot b_2^2$&  0  &0 & -0.90 & -0.90  & - \\
\hline
\end{tabular}
\label{t2}
\caption{ Tree level contributions to the $\pi\pi$ scattering lengths and
slope parameters from the EChL at different orders. The constants of the
 ${\cal L}^{(6)}$ are extracted from dual resonance models as explained in
 the text (see also Table~I)}
  \end{table}

\begin{table}
\begin{tabular}{|r|c|c|c|c|c|}
\hline
    & ${\cal L}^{(2)}$&  ${\cal L}^{(4)}$
& ${\cal L}^{(6)}$ & data from \cite{East} &
 all data\\
\hline
$t_{00}^+$&  0  & 0 & 0& $1.31 \pm 1.26 $  & $0.52 \pm 2.03$ \\
\hline
$t_{01}^+$&  0.63  & 0 & 0& $0.75 \pm 0.09 $  & $0.55 \pm 0.07$ \\
\hline
$10^3 \cdot t_{02}^+$&  0  & 7.2 & -1.23 & -  & - \\
\hline
$10^3 \cdot t_{03}^+$&  0  & 0 & 0.54& -  & - \\
\hline
$10^3 \cdot t_{20}^+$&  0  & 7.2 & 0.62& $17.7 \pm 1.3 $ & $10.3 \pm 1.0$\\
\hline
$10^3 \cdot t_{21}^+$&  0  & 0 & 0.089& $(0.9 \pm 0.15) $ & $(0.2 \pm
0.1)$\\
\hline
$t_{10}^-$&  0.63  & 0 & 0& $1.00 \pm 0.07 $  & $0.52 \pm 0.07$ \\
\hline
$10^3 \cdot t_{11}^-$&  0  & 14.5 &-0.61& $22 \pm 5 $  & $14.9 \pm
2.8$ \\
\hline
$10^3 \cdot t_{12}^-$&  0  & 0 &0.34& $0.33 \pm 0.01 $  & $0.55 \pm
0.1$ \\
\hline
$10^3 \cdot 3t_{30}^-$&  0  & 0 &0.34& - & - \\
\hline
\end{tabular}
\label{t3}
\caption{ Tree level contributions to the low-energy parameters
(defined by eqs.~(3.11, 3.12)) of the
$\pi K$ scattering amplitude
from the EChL at different orders. The constants of the
 ${\cal L}^{(6)}$ are extracted from dual resonance models as explained in
 the text (see also Table~I). The experimental values are the
 results of dispersion-theoretical analysis of Lang and Porod
  [42] }
 \end{table}


\begin{thebibliography}{99}
\bibitem{WeiOld} S.Weinberg,
 Phys. Rev.  {\bf 166}(1968) 1568
\bibitem{Wei} S.Weinberg, Physica  {\bf A96} (1979) 327
\bibitem{GL} J. Gasser, H. Leutwyller, Ann.
 Phys.(N.Y.) {\bf 158} (1984) 142
\bibitem{GLsu3} J. Gasser, H. Leutwyller, Nucl. Phys.
  {\bf B250} (1985) 465
 \bibitem{feasch}
H.W. Fearing and S.Scherer, {\it Extension of the Chiral
Perturbation Theory Meson Lagrangian to $O(p^6)$},
TRI-PP-94-68, [{\tt hep-ph/9408346}]
\bibitem{EckGasPicRaf}
 G.~Ecker, J~.Gasser, A.~Pich and E.~de~Rafael,
  Nucl. Phys. {\bf B321 }  (1989) 311
\bibitem{DonRamVal}
J.~Donoghue, C.~Ramirez and G.~Valencia,
  Phys. Rev. {\bf D39} (1989) 425
 \bibitem{BolPolVer} A. Bolokhov, V.~Vereshagin,  M.~Polyakov,
Sov. J.  of Nucl. Phys., {\bf 53} (1991) 251
\bibitem{BolManPolVer}
 A.~Bolokhov , A.~Manashov,M.~Polyakov and V.~Vereshagin,
  Phys. Rev. {\bf D48} (1993) 3090
\bibitem{Weinber_alg_real} S.~Weinberg,
  Phys.  Rev. {\bf 177} (1968) 2604
\bibitem{Weinberg_mended_symmetry}
S.~Weinberg,
  Phys.  Rev. Lett. {\bf 65} (1990) 1177
\bibitem{DolHorSch}
R.~Dolen, D.~Horn and C.~Schmidt
 Phys. Rev. Lett. {\bf 19} (1967) 402
\bibitem{Veneziano}
G.~Veneziano,
Nuovo Cim.{\bf 57A} (1968) 190
\bibitem{Frampton}
Summary of phenomenological applications of dual models see in \\
P.~Frampton, {\it Dual resonance models}, Benjamin/Cummings,
Menlo Park, CA, (1974)
\bibitem{Polyakov}
A.M.~Polyakov,
{\it Gauge Fields and Strings}, Harwood, (1987)
\bibitem{VVV} V.V.~Vereshagin, Nucl. Phys. {\bf B55} (1973) 621
 \bibitem{BerKaiMei}
 V.~Bernard, N.~Kaiser and U.-G.~Mei\ss ner,
  Nucl. Phys. {\bf B357 } (1991) 129
\bibitem{Col}
G.~Colangelo,
  Phys. Lett., {\bf 350B} (1995) 85
  \bibitem{KneMouSteFuch}
 M.~Knecht, B.~Moussallam, J.~Stern and N.H.~Fuchs,
 {\it The Low Energy $\pi\pi$ Amplitude to One and Two Loops},
 IPNO/TH~95-45, PURD-TH-95-05, [{\tt hep-ph/9507319}]
\bibitem{two}
J.~Gasser, { \it private communication }
\bibitem{Lew}
D.~C.~Lewellen,
  Nucl. Phys. {\bf B392} (1993) 137
\bibitem{Centr}
J.-R.~Cudell, K.R:~Dienes,
Phys. Rev. Lett. ,{\bf 69} (1992) 1324
\bibitem{mylec}
M.~V.~Polyakov,
{\it Effective Chiral Lagrangians versus Duality},
lectures given at the Third St.~Petersburg Winter School, (1995),
to be published
\bibitem{DiaEid} D.I. Diakonov, M.I. Eides,
JETP Lett. {\bf 38} (1983) 433
\bibitem{Bal} J.Balog,
  Phys. Lett., {\bf 149B} (1984) 197
\bibitem{And} A.A.~Andrianov, Phys. Lett., {\bf 157B} (1985) 425
 \bibitem{DiaPet} D.I.~Diakonov, V.Yu.~Petrov,
  Nucl. Phys. {\bf B272} (1986) 457
\bibitem{Regge_review}
 P.D.B.~Collins and A.D.~Martin,
 {\it Hadron interactions}, Adam Hilger, Bristol, (1984)
\bibitem{tHooft74}
 G.~'t~Hooft,
  Nucl. Phys. {\bf B72 } (1974) 461
\bibitem{Lovelace}
C.~Lovelace,
  Phys. Lett., {\bf 28B} (1968) 264
\bibitem{Shapiro}
J.A.~Shapiro,
  Phys.  Rev. {\bf 179} (1969) 1345
\bibitem{Migdal_Makeenko}
For a review see: A.A.~Migdal, Phys. Rep.{\bf 102} (1983) 199
\bibitem{lattice}
K.G.~Wilson,
  Phys.  Rev. {\bf D10} (1974) 2445
\bibitem{Gross}
D.~Gross and W.~Taylor,
  Nucl. Phys. {\bf B400 } (1993) 181
\bibitem{KawKitYab}
K. Kawarabayashi, S. Kitakado and H.Yabuki,
Phys. Lett., {\bf 28B} (1969) 432
\bibitem{BijColGas}
J.~Bijnens, G.~Colangelo and J.~Gasser,
  Nucl. Phys. {\bf B427 } (1994) 427
\bibitem{Slava}
V.~A.~ Kudryavtsev,
Yad. Phys., {\bf 58} (1995) 137
\bibitem{ManGeo}
A.~Manohar and H.~Georgi,
  Nucl. Phys. {\bf B234 } (1984) 189
\bibitem{Zuk}
J.~Zuk,
Z.~Physik, {\bf 29} (1985) 303
\bibitem{PraVal}
M.~Praszalowicz and G.~Valencia,
  Nucl. Phys. {\bf B341 } (1990) 27
\bibitem{Nag}
M.M. Nagels et al.,
  Nucl. Phys. {\bf 147 } (1979) 189
  \bibitem{LanPor}
  C.B.~Lang and W.~Porod,
  Phys.  Rev. {\bf D21} (1980) 1295
  \bibitem{East}
  P.~Estabrooks et al.,
  Nucl. Phys. {\bf B133} (1978) 490
\bibitem{hep-echl6}
 A.~Bolokhov , A.~Manashov,M.~Polyakov and V.~Vereshagin,
{\it Contribution of the Sixth Order Chiral Lagrangian to the $\pi K$
Scattering at Large $N_c$},[{\tt hep-ph/9503424}]
\end{thebibliography}
\end{document}